%% file: eCrime-main.tex
\documentclass[10pt,conference]{IEEEtran}

\IEEEoverridecommandlockouts

\usepackage{xspace}
\usepackage{graphicx}
\usepackage{subfig}
\usepackage{float}
\usepackage{multirow}
\usepackage{comment}
\usepackage{booktabs}
\usepackage[hidelinks]{hyperref}
\newcommand{\name}{{\scshape EagleEye}\xspace}

\title{\name: Attention to Unveil Malicious Event Sequences from Provenance Graphs}

\begin{document}

\author{
    \IEEEauthorblockN{Philipp Gysel\IEEEauthorrefmark{1},
                    Candid Wüest\IEEEauthorrefmark{1},
                    Kenneth Nwafor\IEEEauthorrefmark{1}\IEEEauthorrefmark{2},
                    Otakar Jašek\IEEEauthorrefmark{1},
                    Andrey Ustyu\-zhanin\IEEEauthorrefmark{3}\IEEEauthorrefmark{1},
                    Dinil Mon Divakaran\IEEEauthorrefmark{4}}
    \IEEEauthorblockA{
        \IEEEauthorrefmark{1}Acronis Research,
        \IEEEauthorrefmark{2}Constructor Technology, \\
        \IEEEauthorrefmark{3}Constructor University, Bremen,
        \IEEEauthorrefmark{4}A*STAR Institute for Infocomm Research
    }

    \thanks{Corresponding author: Philipp Gysel (gyselph@gmail.com). This work was done when all authors were affiliated with Acronis Research.}
}

\maketitle

\begin{abstract}

Securing endpoints is challenging due to the evolving nature of threats and attacks. With endpoint logging systems becoming mature, provenance-graph representations enable the creation of sophisticated behavior rules. However, adapting to the pace of emerging attacks is not scalable with rules. This led to the development of ML models capable of learning from endpoint logs. 
However, there are still open challenges:~i)~malicious patterns of malware are spread across long sequences of events, and ii)~ML classification results are not interpretable. 

To address these issues, we develop and present \name, a novel system that 
i)~uses rich features from provenance graphs for behavior event representation, including command-line embeddings, ii)~extracts long sequences of events and learns event embeddings, and iii)~trains a lightweight Transformer model to classify behavior sequences as malicious or not. We evaluate and compare \name against state-of-the-art baselines on two datasets, namely a new real-world dataset from a corporate environment, and the public DARPA dataset.
On the DARPA dataset, at a false-positive rate of 1\%, \name detects $\approx$89\% of all malicious behavior, outperforming two state-of-the-art solutions by an absolute margin of 38.5\%.
Furthermore, we show that the Transformer's attention mechanism can be leveraged to highlight the most suspicious events in a long sequence, thereby providing 
interpretation of malware alerts.

\end{abstract}


\input{intro}
\input{motivation}
\input{threat-model}
\input{arch}
\input{behavior-modeling}
\input{perf-eval}
\input{case-studies}
\input{related}
\input{conclusions}

\bibliographystyle{IEEEtran}
\bibliography{references}

\input{appendix}

\end{document}

%% file: intro.tex
\section{Introduction}
\label{sec:introduction}

The rate of cyber attacks, their sophistication, the number of targets, and the losses the attacks cause, have clearly witnessed a worrying upward trend over time. In May 2024 alone, more than 9~million new malware samples were reported~\cite{avatlas}. 
Reports estimate that cybercrime will cost the world more than 9~trillion USD in 2024~\cite{cybercrimecosts}.
With a mature underground market, cybercriminals have
easy access to new, better, and advanced malware to assist them with their nefarious goals. 

Detection of malware infection is challenging. 
With wide adoption of TLS, much of the network communication 
is encrypted today, thus limiting the capabilities of perimeter defense systems such as firewalls, deep packet inspection, and intrusion detection systems~\cite{disclosure-2012,TLS-based-ML-2017, evidence-gathering-2017, NADA-2018, GEE-CNS-2019}.
This limitation is turning the point of observation onto endpoints, where we have the capability to monitor and log low-level kernel calls of all activities,
including process creations, file system activities, and network connections.
EDR (endpoint detection and response) solutions such as CrowdStrike Falcon~\cite{crowdstrike-falcon}, SentinelOne Singularity~\cite{sentinelone-singularity}, and Trend Micro Apex One~\cite{trend-apex-one} can log such behavior. 
These logs present a
rich source of information useful for
security auditing~\cite{OmegaLog-NDSS-2020}. Hence, despite the computational and storage costs involved, security vendors have been developing and enhancing EDR solutions that provide high visibility into activities happening
on endpoints via continuous logging. Importantly, EDR solutions are complemented with signatures to detect known malware, which are updated continuously as new malware samples are discovered. 

Yet, simple rules that, say, match against hashes of known malicious binaries and IP addresses with bad reputation, are not sufficient to counter emerging threats. Sophisticated attacks exploit or disguise as benign applications, encrypt individual binary segments, employ polymorphic techniques, obfuscate network communication to exfiltrate sensitive information, and use other evasion techniques.
This observation led to the development of {\em provenance graphs} as a dependency  structure depicting entities  (such as processes, files, etc.) at endpoints and their relationships. With such graphs, a security analyst can
visually see the relationship between system entities
and identify suspicious patterns~\cite{PG-EDR-CCS-2023}. 
These graphs also facilitate the writing of more sophisticated signatures which can be used to detect malware behavior spread across a provenance graph (see Section~\ref{sec:why_features} for an illustration).

While signatures crafted by security analysts are precise in detecting known malware, there are important disadvantages. i)~Analyzing graphs and manually writing rules is not a scalable approach, given that there are more than 300,000 new malware samples detected
daily~\cite{avatlas}. ii)~Moreover, 
the provenance graphs are themselves huge in size~\cite{sok-provenance-2023}. iii)~Attacks keep evolving with time, and with attackers now getting powerful AI tools to assist them in the process (e.g., FraudGPT~\cite{fraudGPT}), the quantity and quality of attacks are only expected to rise, thus compounding the challenge. Consequently, researchers have been developing machine learning models for investigating incidents, triaging, and detecting malware~\cite{deeplog-2017,tiresias-2018, alsaheel2021atlas, holmes-2019, NoDoze-2019,han2020unicorn,shadewatcher-2022,van2022deepcase, wang2020you,ProGrapher-2023}.

The following challenges are yet to be addressed. 
The malicious behavior of a malware sample can be spread out in a (potentially) long sequence of events in a provenance graph; existing works use models that are not able to effectively learn from such long sequences of events. This results in low detection rates.
Second, it is not only important to classify behavior as malicious or not, but also to provide interpretation of results, so that an analyst can efficiently assess the criticality of an incident. Addressing these challenges, we develop \name, 
a novel system that 
i)~uses rich features from behavior events of applications, including command-line strings, ii)~extracts and embeds long sequences of events, and
iii)~employs a lightweight Transformer model that learns to predict the intent of an application.
We summarize our contributions:

\vspace{-0.1cm}

\begin{itemize}
    \item We develop a novel system called \name, which processes low-level events and learns the behavior of applications.
    We explore the capability of Transformers in modeling endpoint behavior based on process provenance graphs (Section~\ref{sec:arch}). 
    
    \item We enrich  a provenance graph with features capturing the context of behavior events (Section~\ref{sec:enrich_graphs}).
    Different from previous works, \name employs another Transformer to generate embeddings of command-line strings used to start new processes, thus 
    capturing command details and hierarchy in the process tree (Section~\ref{sec:cmdline_embedding}). 
    
    \item We perform a systematic evaluation and compare \name with state-of-the-art solutions, namely, Deep\-CASE~\cite{van2022deepcase} and ProvDetector~\cite{wang2020you}.
    Our experiments on two datasets demonstrate that, in comparison to the baselines, \name achieves significantly higher detection accuracy at very low false positive rates.
    
    \item We evaluate the modeling capability of the Transformer model in \name and compare it with other ML models (Section~\ref{subsubsec:transformer-model-results}), such as LSTMs used in existing works~\cite{deeplog-2017, alsaheel2021atlas, villarreal2021hunting}. We also study the impact of command-line embeddings and security features in achieving high performance
    (Section~\ref{subse:command-line-results}).
     
\end{itemize}

Moreover, we carry out a case study, which reveals that \name can both learn the malicious context from a long sequence and provide an explanation for the identified malicious pattern (Section~\ref{sec:case-study}).
To support further research, we publish the code of \name, along with the new malware dataset we generated\footnote{Implementation and malware dataset: \url{https://github.com/gyselph/eagle-eye}}.

%% file: motivation.tex
\section{Importance of feature-rich provenance graphs}
\label{sec:why_features}

Provenance graphs are a powerful tool to differentiate between malicious and benign applications. In this section, we demonstrate the importance of
encoding detailed behavior information into the graph nodes. For this purpose, we proceed to analyze a concrete malware type and its behavior, which provides insights into how a malware detection system can identify the intent of an application.
During this process, we show how a professional security analyst would craft a behavior signature for this specific malware.

\begin{figure}[]
  \centering
  \includegraphics[width=\linewidth, clip, trim=3.7cm 7.2cm 13cm 7.7cm]{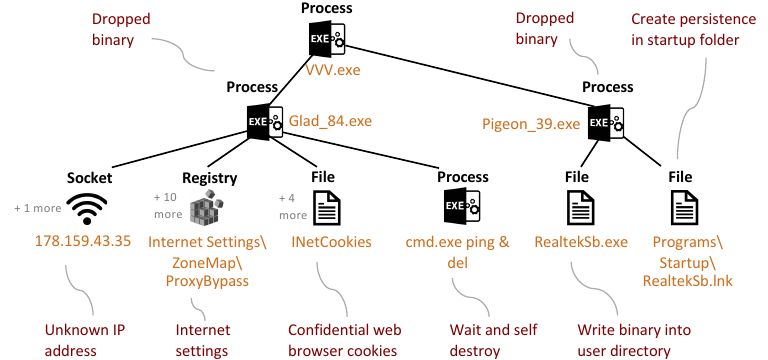}
  \caption{Provenance graph of Raccoon Infostealer malware}
  \label{fig:raccoon}
  \vspace{-0.3cm}
\end{figure}

Figure~\ref{fig:raccoon} shows the most relevant parts of the provenance graph of the well-known Raccoon Infostealer~\cite{raccoon-darktrace}. We downloaded a sample instance from VirusTotal~\cite{virus2021}, executed it
in a controlled environment, and tracked its runtime behavior using a commercially available EDR tool. In what follows, we describe its behavior according to the MITRE ATT\&CK framework~\cite{mitre-attack}. After initial access of executable \texttt{VVV.exe}, an application \texttt{Glad\_84.exe} is written to disk and started. The application first reaches out to a C\&C  server. In order to detect if it is within a sandbox environment, the application checks various registry keys. As part of the collection phase, the application reads multiple web browser cookies, which potentially contain confidential information. Finally, to evade detection, it executes \texttt{cmd.exe} with appropriate flags to delete the binary.
To ensure that this indicator removal happens \textit{after} the executable has completed its work, a prolonged \texttt{ping} operation is prepended to the delete step.
In parallel, a second application \texttt{Pigeon\_39.exe} is also written to disk and started. To gain persistence, a link file is written to the startup directory. This link file points to a dropped binary \texttt{RealtekSb.exe}, which is written to the user directory for masquerading purposes.

Security vendors typically create behavior signatures manually. For the studied malware sample, such a rule could be summarized as follows: i)~The provenance graph contains at least one dropped binary that is executed. ii)~A connection is established to a new and potentially suspicious IP address. 
iii)~Multiple registry keys related to host settings are read. iv)~Multiple sensitive files are read. 
v)~Persistence is created via a dropped binary in the startup directory.
vi)~The main file deletes itself through a ping sleep \& delete command.

As one can observe, this behavior signature uses both global information as well as fine-grained details of individual actions. For example, condition 1 above requires global knowledge spread across multiple nodes of the provenance graph, to check if a started executable was previously persisted
in the same graph. On the other hand, condition~3 can only be checked if the exact registry key path is known. Finally, conditions~4 and~5 above mandate that the precise file path for each file event is given.

\begin{figure}[]
  \centering
  \includegraphics[width=\linewidth, clip, trim=4cm 7.8cm 13.5cm 7.7cm]{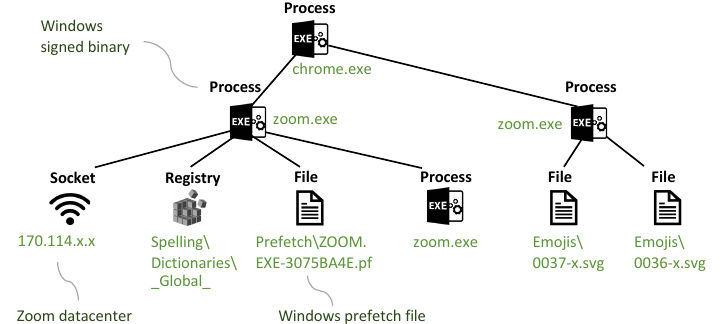}
  \caption{Provenance graph of a Chrome instance}
  \label{fig:benign_chrome}
  \vspace{-0.2cm}
\end{figure}

Next, we switch our focus to a benign provenance graph from a \texttt{Chrome.exe} instance. 
Figure~\ref{fig:benign_chrome} shows its key behavior, which we captured on an endpoint under real-world conditions, using the same EDR tool as for the malware sample. As shown in the figure, the web browser starts two independent Zoom sessions. The first session performs a short query to a Zoom data center, checks the local language dictionary, uses a prefetch file for faster execution, and starts a nested Zoom session. The second session is only used for chatting, and emojis are used during typing.

With detailed information present in each behavior node, both the benign and malicious provenance graphs can be easily categorized by a security expert. Unfortunately, most existing solutions drop the rich information from the graph nodes and only keep the entity types (e.g., \texttt{file}, \texttt{process}, \texttt{socket}, etc.) and associated actions (e.g., \texttt{read}, \texttt{write}, \texttt{execute}, etc). As a case in point, let us consider the final file creation of both provenance graphs. With the help of all event details, the last file creation of the Raccoon Infostealer is potentially suspicious, as a link file is created in the Autostart folder, pointing to a dropped binary. However, if this event was only represented by the entity type (\texttt{file}) and the associated action (\texttt{write}), this event would look identical for both provenance graphs, thus opening the door for evasion by attackers~\cite{goyal2023sometimes}.

In fact, most state-of-the-art proposals for malware detection use coarse behavior features. ProGrapher~\cite{ProGrapher-2023} and Unicorn~\cite{han2020unicorn} consider only the entity and action type. As already mentioned, this approach causes the last file event of both aforementioned provenance graphs to look identical.
Similarly, DeepCASE~\cite{van2022deepcase} uses high-level SIEM events, where each event is represented by one categorical feature. By contrast, in our work, we propose to use a large feature vector per behavior event, of size 60, where the feature vector contains rich information typically used by experienced security analysts (see Section~\ref{sec:enrich_graphs}). We argue that this additional and relevant information makes it possible to clearly differentiate graphs like those in Figures~\ref{fig:raccoon} and~\ref{fig:benign_chrome} into malicious and benign ones.

We conclude that tracking low-level kernel calls of an application, and keeping detailed features about each such behavior event, allow a human analyst to classify applications into benign and malicious. Therefore, we provide the same behavior data to \name, with the goal of detecting malicious behavior on endpoints. 

\vspace{-0.05cm}

%% file: threat-model.tex
\section{Threat model}

Our threat model is similar to previous works which use endpoint log analysis for malware detection~\cite{OmegaLog-NDSS-2020, wang2020you, shadewatcher-2022,threatrace-2022, van2022deepcase,ProGrapher-2023}. We assume a system in place that monitors kernel-level calls, such as registry events, network connections, file creations, etc. Commercially available EDR tools are capable of logging such events for security purposes. 

We consider attacks on endpoint computers achieved by malware. We refer to the provenance graph corresponding to a malware's action on an endpoint as a malicious graph. Thus, the goal of our work is to perform online detection of malware on endpoints.
A malware sample might be of any family or variant, and as such might have different goals, such as infecting other machines, or exfiltrating sensitive data.
During this process, the malware sample may also be involved in other activities. 
In particular, many malware variants are known to disguise their intention by instrumenting legitimate processes to achieve their goals. This so-called Living-off-the-land technique~\cite{ongun2021living,barr2021survivalism} can be used to inject code into an existing legitimate system process, or to start a new legitimate process which performs part of the malicious work.
Despite this obfuscation, the kernel-level driver sees the API requests performed by the legitimate process
and logs such information. Similar to past works, we assume that the auditing framework and log files are not compromised, via the use of a trusted computing base and hardening techniques.

%% file: arch.tex
\section{\name: Architecture and model}
\label{sec:arch}

\begin{figure*}[]
  \centering
  \includegraphics[clip, trim=2cm 7.5cm 10.6cm 5.9cm, width=12.3cm]{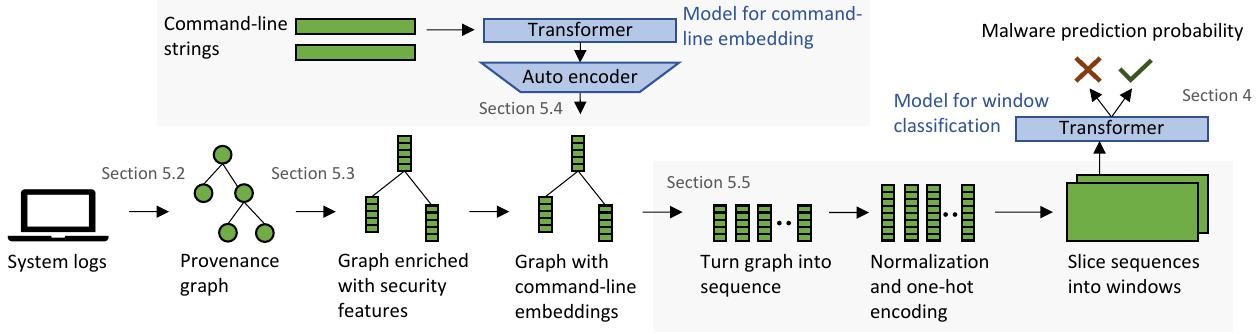}
  \caption{System architecture of \name}
  \label{fig:system_diagram}
  \vspace{-0.1cm}
\end{figure*}

Figure~\ref{fig:system_diagram} presents an overview of \name. The left side of the figure illustrates the feature extraction mechanism and the process of transforming system logs into a data format suitable for ML modeling. As shown on the right side of the figure, the enriched and preprocessed data is then fed into the trained model for malware classification. 
Next, we present the modeling aspects of \name. 

\subsection{Transformer for learning behavior events}

Since the seminal work by Vaswani et al.~\cite{vaswani2017attention}, Transformers have demonstrated their capability in various natural language processing (NLP) tasks. The Transformer architecture, which is based on the self-attention mechanisms, has shown significant efficacy, surpassing the previous state-of-the-art  models such as recurrent neural networks (RNNs), making it the de facto standard in NLP today.
However, past works on malware detection use RNNs~\cite{tiresias-2018, van2022deepcase,villarreal2021hunting,alsaheel2021atlas,deeplog-2017}; we argue that Transformer models exhibit a significantly better capability for classifying computer applications, as also confirmed by our experiments (Section~\ref{subsubsec:transformer-model-results}).

{\it Problem of long sequences}: Malware detection is akin to the problem of finding a needle in a haystack. In the example of the Raccoon Infostealer from Figure~\ref{fig:raccoon}, only a handful of events reveal its malicious intent. However, over the lifespan of the Raccoon executable, more than 1,000 behavior events are observed. Therefore, a significant part of the behavior sequence is irrelevant for malware detection. Transformers can deal much better with long sequences than RNNs, thanks to their attention mechanism~\cite{vaswani2017attention}. 

{\it Problem of complex context}: 
Application classification requires detection of complex patterns, which in turn depends on  multiple correlated events that are spread far apart in a sequence. 
Revisiting the
Raccoon Infostealer (Figure~\ref{fig:raccoon}), it is the \textit{combination} of C\&C communication, reading of sensitive files, persistence creation, and final uploading of data that turns the behavior into a malicious pattern. If 1-2 behavior events are missing, 
the pattern may be categorized differently and as benign.
Transformers are able to extract such complex patterns, as 
the internal attention mechanism 
attends to the whole input sequence for context creation. 
As we show in Section~\ref{sec:case-study}, \name's model indeed  pays high attention to suspicious events while ignoring unimportant ones.

\subsection{Embedding behavior events}
\label{subsec:embedding-events}

\begin{figure}[]
  \centering
  \includegraphics[clip, trim=3.2cm 6.1cm 17.0cm 5.1cm, width=6.5cm]{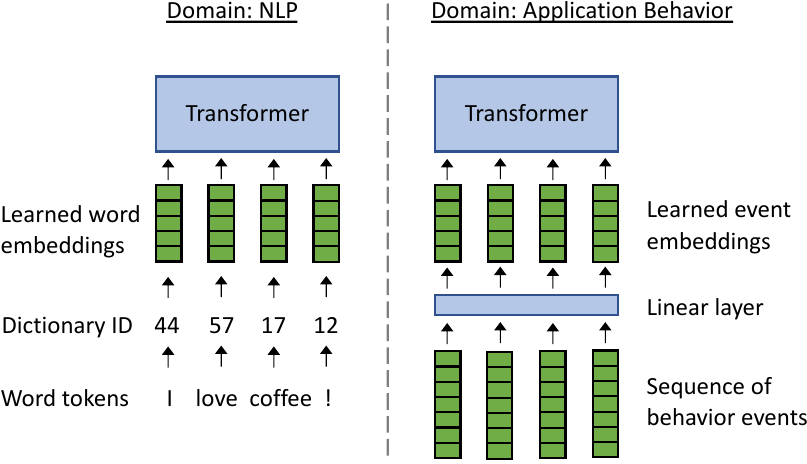}
  \caption{Token embedding: NLP tasks vs. \name}
  \label{fig:token_embedding}
  \vspace{-0.1cm}
\end{figure}

\name represents a behavior event as a fixed-size vector of security features. Figure~\ref{fig:token_embedding} illustrates our event embedding approach (right), and compares it with the traditional word embedding in the NLP domain (left).
In the NLP setting, an embedding layer translates token IDs into word embeddings, which are learned during  the training phase. In our setting, each behavior event is represented as a fixed-size vector.
Our architecture contains a linear embedding layer, which learns (via back-propagation) to project the behavior events into a new latent feature representation. With this embedding layer, we reduce the dimension of each behavior vector from \(\sim \)200 to an embedding vector of dimension 60, 
effectively reducing the sparse high-dimensional space to a smaller latent space. Next, we describe the architecture of the proposed Transformer model. 

\subsection{Transformer architecture}

There exist three Transformer architecture types: \textit{encoder-only}, encoder \textit{and} decoder, and \textit{decoder-only} architecture. The first type is suited for classification problems, whereas the second and third are more complex architectures used for generative tasks. Since malware detection is a classification task, we choose to use an encoder-only architecture.
Our goal is to build a lightweight malware classification system, which mandates a small-sized resource-efficient model. Therefore, we build a small Transformer model, which is based on BERT-Tiny~\cite{turc2019well}.

\begin{figure}[]
  \centering
  \includegraphics[clip, trim=7.5cm 6.4cm 14.6cm 4.6cm, width=6.5cm]{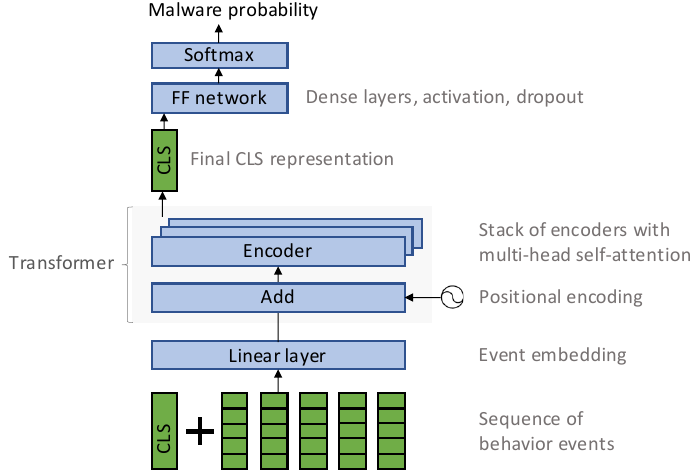}
  \caption{Transformer architecture of \name}
  \label{fig:transformer_architecture}
  \vspace{-0.1cm}
\end{figure}

We present the overall architecture of \name's Transformer model in Figure~\ref{fig:transformer_architecture}. The model input consists of an event sequence, where each event is represented by security features (as described in the next section). Similar to the original BERT model, we prepend each sequence with a \texttt{CLS} token. The event sequence plus \texttt{CLS} token are passed through a linear projection layer for event embedding.
The positional encoding in \name is based on the event sequence index. The resulting sequence goes through an encoder stack with multi-head self-attention. In the final encoder layer, only the \texttt{CLS} token's representation is used, and all other tokens 
are discarded. The \texttt{CLS} token representation is passed through a 2-layer dense network, followed by a softmax layer, which produces the final malware classification probability.

%% file: behavior-modeling.tex
\section{\name: Behavior modeling}
\label{sec:behavior-modeling}

\subsection{Overview of data processing pipeline}

\begin{figure}[]
  \centering
  \includegraphics[width=8cm, trim=2cm 5.6cm 11cm 6.5cm, clip]{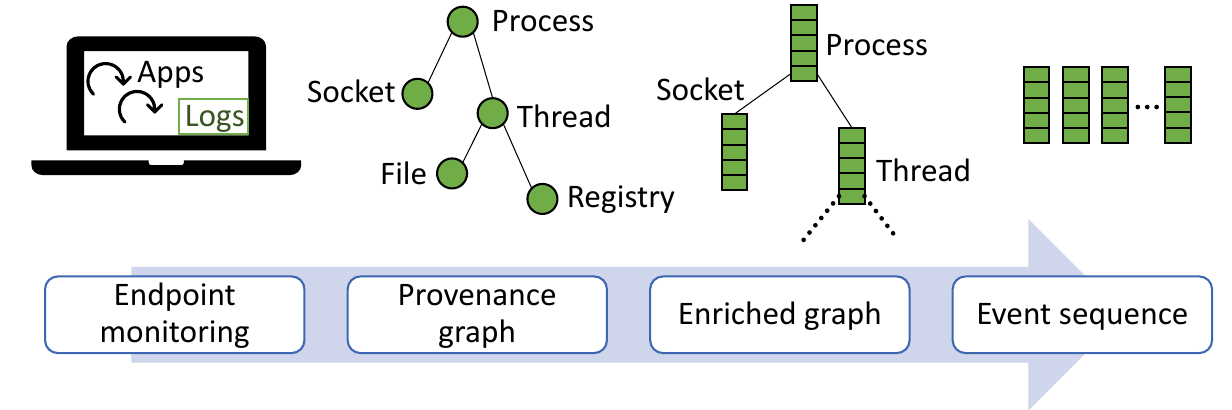 }
  \caption{Application behavior modeling}
  \label{fig:sequence_modeling}
\end{figure}

Figure~\ref{fig:sequence_modeling} illustrates the transformation of security logs into graph-based sequences, which serve as input to \name.

 1) First, we deploy a system that tracks all low-level kernel calls of running applications on endpoints.     To avoid an overload of data, we apply a filtering mechanism, and only collect a subset of all incoming event types (Section~\ref{sec:logs_to_graphs}).
    
 2) We convert security logs into provenance graphs, which represent behavior events via raw features (Section~\ref{sec:logs_to_graphs}). 
    
3.1) We enrich a provenance graph with security relevant features (Section \ref{sec:enrich_graphs}). During this process, we go beyond raw features and extract  advanced features that learn context from other parts of the provenance graph. 

 3.2) A novel aspect of \name is the inclusion of command-line embeddings, which represent commands and their hierarchy in the process tree (Section~\ref{sec:cmdline_embedding}).
    
 4) Finally, a graph gets converted to a sequence of chronologically ordered nodes (Section~\ref{sec:graphs_to_windows}). As each node is represented by a fixed-size vector of security relevant features, the application behavior is now modeled via a sequence.

\subsection{Convert a system log to provenance graphs} \label{sec:logs_to_graphs}

Under real-world conditions, user endpoints host a variety of applications, where each application typically produces hundreds of behavior events every minute.
Similar to previous works~\cite{alsaheel2021atlas,wang2020you,threatrace-2022, ProGrapher-2023}, we use provenance graphs as an intermediate data representation. \name builds provenance graphs by creating a node for each behavior event
and edges for capturing the parent-child relationships between the different events.
We represent provenance graphs as follows:
we store all behavior information in nodes, and edges do not contain any information. Nodes contain information about both the action taken (e.g. read, write, execute) as well as
the involved system entity (e.g. file name, path, file access mode, etc). A provenance graph's head node represents the bootstrapping process. All its child nodes are actions taken directly by this process, including the creation of new processes. Child processes, in turn, will point to new nodes containing actions initiated by them.
Clearly, the provenance graph thus created is a DAG (directed acyclic graph). 

EDR solutions gather various details for each behavior event, beyond just process names and call hierarchies. We use only a subset of all available information, which we store as \textit{raw} features in nodes of the corresponding graph. Such raw features can be simple in nature, like file names and process names, which are also used in other works~\cite{wang2020you}. We also store other particulars like file access flags and  command-line flags.

Every operating system has its own bootstrapping processes, which are involved during startup, and later used for orchestration of compute loads. On Windows OS, user processes are started directly or indirectly by a handful of bootstrapping processes, like \texttt{winlogon.exe}, \texttt{win\-init.exe}, and \texttt{logonui.exe}. Since different applications like \texttt{note\-pad.exe} and \texttt{outlook.exe} should be analyzed separately, we do not include the bootstrapping processes as a common ancestor, and instead create disjoint provenance graphs for the user applications. 

To construct provenance graphs from a system log, we select a random behavior event, and iteratively move up the parent-child hierarchy until we reach a native OS process. This corresponding child node (with a native OS process as parent) serves as the head of the provenance graph.
The head node will only have outgoing edges and no incoming ones. We now perform breadth-first search from the graph head. For each process, we add all child processes as child nodes to the graph. Moreover, for each process, we search for all initiated kernel calls, like file accesses and network connections, and add these events as child nodes to the corresponding process node.
The resulting provenance graph contains a subset of all behavior events from the system log, as there are multiple disjoint applications running concurrently on a given system.

Next, we pick another behavior event which did not get added to the previous set of provenance graphs, and proceed to create another provenance graph in the same manner. We continue building provenance graphs until each behavior event is either represented in one of the graphs, or was initiated from the operating system and is thus ignored. 
Thanks to this grouping of data into different graphs, we achieve separation of data, which can then be analyzed independently. Each graph gives an account of where the application interactions originate from. In the example of the Raccoon Infostealer (Figure~\ref{fig:raccoon}), the provenance graph  explains where the malicious persistence \texttt{RealtekSb.lnk} comes from, and thus the head process \texttt{VVV.exe} from the graph can be flagged as suspicious.

Our data processing pipeline has several built-in measures to control the size of the resulting graphs. First, we do not include processes from the operating system. Next, we do not include all existing behavior event types, but only 
the most relevant ones from a security perspective.
Table~\ref{tab:behavior_events} gives a list of behavior events captured by most EDR systems; however we limit \name to only use four event types (see Table~\ref{tab:sec_features} in the Appendix). Finally, we set a maximum duration for each provenance graph, and start a new graph after the timer expires.
While a longer duration helps to capture more dependencies and relationships among the different events, the resulting larger graph also creates computational and storage challenges. Furthermore, the goal of \name is to detect
malicious software as early as possible, before 
any harm is inflicted on the endpoint.

\begin{table}[]
\caption{Behavior event types collected by EDR solutions}\label{tab:behavior_events}
\vspace{-0.1cm}
\small{
\begin{center}
\begin{tabular}{lll}
\toprule
\textbf{Processes/threads} & \textbf{Miscellaneous} & \textbf{Suspicious} \\
\midrule
Process start & Socket & Stack pivot \\
Process termination & File access & Memory protection \\
Thread creation &  File permission & Hook injection \\
Thread suspension & Registry key & Process hollowing \\
Thread resumption & RPC call & Token manipulation \\
\bottomrule
\end{tabular}
\end{center}
}
\end{table}

\subsection{Graph enrichment with security features}
\label{sec:enrich_graphs}

There are two main goals during the graph enrichment phase. i)~Add more context-aware information to each graph node, which will provide \name with relevant information for malware detection. ii)~The \textit{raw} features of graph nodes should be turned into preprocessed \textit{security} features,
which are in a structured format suitable for machine learning.

During the enrichment phase, \name leverages the context of the whole graph to condense global information into features of individual nodes. As an example, the process call hierarchy may reveal important details of an application's intent~\cite{ongun2021living, OmegaLog-NDSS-2020}. We therefore use the process hierarchy and each process' command and flags to create call-chain aware command-line embeddings (Section~\ref{sec:cmdline_embedding}). As another example, consider \textit{dropped binaries}, i.e., an executable  {\it recently} downloaded from the Internet, written to disk, and then started as a new process. We define a boolean feature for process nodes to indicate whether an executable was dropped.

Each behavior event type has its own set of \textit{raw} features. These raw features are turned into structured \textit{security} features. A full list of our 24 security features is included in Appendix \ref{sec:appendix_security_features}, Table~\ref{tab:sec_features}. As a case in point,
for each registry access, we propose a security feature which indicates whether a particular registry key is creating persistence for an application. To achieve this, we leverage the full registry key path.
This feature simplifies the task for the subsequent  learning model, when compared to using the raw registry key path. As an important observation, there are other registry keys with the opposite effect, such as
notify keys. Notify keys create a graphical notification for the user, and are typically used by benign applications; we represent notifications using a boolean feature.

As another case in point, for representation of file access events, we introduce a security feature which captures whether the file in question contains sensitive information, such as credit card data, passwords, or web cookies.
In addition, we use a categorical security feature that represents the location of a given file. For this purpose, we split the whole file system into coarse groups which share a common semantic meaning. For Windows, one such important path is the autostart directory. Other important path categories include temporary directories, system directories, and user directories.

The graph enrichment step is a novel component of \name, which, to the best of our knowledge, is missing in existing detection solutions. Previous works either do not consider a detailed provenance graph~\cite{villarreal2021hunting, deeplog-2017, van2022deepcase} or use a provenance graph which contains only \textit{raw} features~\cite{alsaheel2021atlas, wang2020you, han2020unicorn, ProGrapher-2023, shadewatcher-2022, han2021sigl, threatrace-2022}.

\subsection{Command-line embedding}
\label{sec:cmdline_embedding}

\begin{figure}[]
  \centering
  \includegraphics[clip, trim=3.6cm 10.3cm 6.2cm 2.8cm, width=8.5cm]{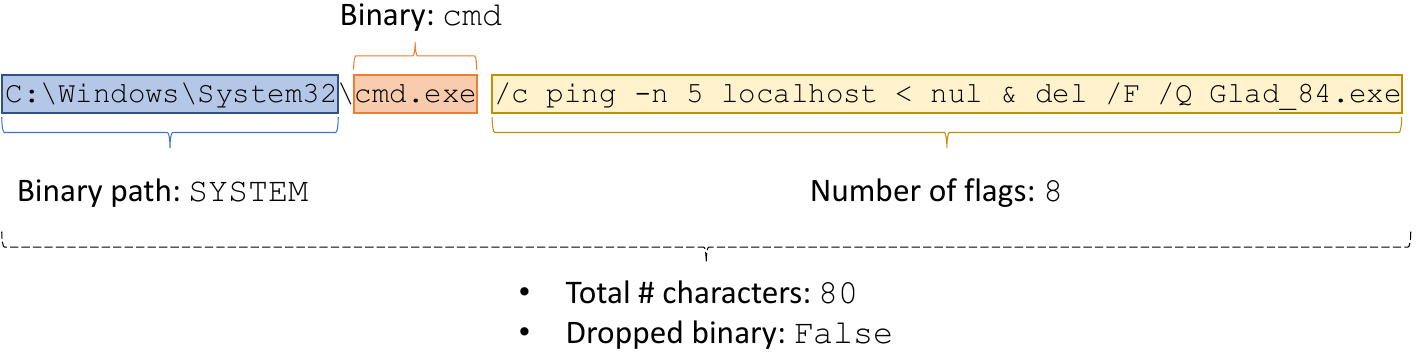}
  \caption{Featurized representation of a command-line string}
  \label{fig:cmdline_representation}
\end{figure}

Process creations, and in particular the command-line strings used for process starts, are relevant for malware detection. In fact, some recent works~\cite{ongun2021living,filar2020problemchild} rely \textit{only} on command-lines to perform malware detection. To demonstrate the importance of command-lines, we investigate the wait-and-self-destroy operation of the Raccoon Infostealer mentioned earlier. The full command-line string with the executable path, the name, and all flags is shown in Figure~\ref{fig:cmdline_representation}. A security analyst will consider this command  as suspicious, since a benign  program would rarely hide a wait operation in a \texttt{ping} command and subsequently delete itself.
The challenge here is to represent this data in a format that is understandable by a  model. Typically such command-line strings are represented by handcrafted features, like the string length, the path, and the name of the binary, etc. However, such features may not capture all relevant information.

Another approach is to represent the whole string using a certain embedding. Filar et~al.~\cite{filar2020problemchild} used the word frequency algorithm TF-IDF for this purpose, and in~\cite{ongun2021living}  word2vec embedding~\cite{mikolov2013efficient} was applied. 
However, the Transformer architecture has demonstrated a superior capability for text reasoning, with models like BERT~\cite{devlin2018bert} being used to generate embeddings for security tasks such as phishing email detection~\cite{d-fence-2021}.

For our representation of commands, \name uses a tiny \texttt{all\_MiniLM\_L6\_v2} Transformer~\cite{all_MiniLM_L6_v2}, which is  pre-trained on a large corpus of data to generate text embeddings.
As input to this sentence Transformer, we use the whole command-line string
of both the current and parent processes. Thus the model captures: i)~richer semantics than what handcrafted features do,  
ii)~the process hierarchy. As output of the Transformer, we compute the mean of all output tokens, resulting in a 384-dimensional vector. To compress this vector into a more manageable size, we train an autoencoder in unsupervised fashion. That is, we pass all command-line strings from the training dataset through the sentence Transformer, and forward the output to the autoencoder, which then learns to compress the command-line representation into a 16-dimensional vector.
The resulting vector serves as an additional security feature for process creations (see Figure~\ref{fig:system_diagram}).

\subsection{Convert graphs into windows}
\label{sec:graphs_to_windows}

\textbf{Graph to sequence}. Once provenance graphs are enriched with security features and command-line embeddings, each graph is turned into a sequence of events. For this purpose, we take a full provenance graph, and traverse its nodes in temporal order, to form a sequence of events. This gives \name the following benefits. One, based on a predefined sequence length, we can process a graph as soon as we have that many events. This allows for early detection of malware.
Two, sequences can be stored and traversed in an efficient way. For a real-world endpoint detection system, low memory consumption is of high importance. A memory-optimized implementation of \name needs to keep only two data structures in memory: a continuously updated tree with the process call hierarchy, and a flat list of recent behavior events.

\begin{figure}[]
  \centering
  \includegraphics[clip, trim=6.3cm 11.6cm 12.9cm 5cm, width=8cm]{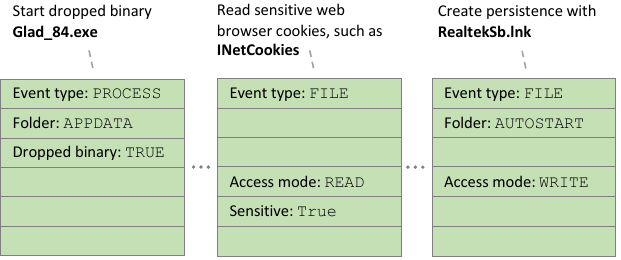}
  \caption{Sequence of events represented by security features, for Raccoon Infostealer}
  \label{fig:behavior_sequence}
  \vspace{-0.5cm}
\end{figure}

When the graph-to-sequence transformation is applied to the provenance graph of the Raccoon Infostealer (from Figure~\ref{fig:raccoon}), we get a sequence of fixed-size vectors as shown in Figure~\ref{fig:behavior_sequence} (for readability, only three important behavior events are shown). 
The resulting sequence consists of security feature vectors, where each vector element is of categorical, numerical, or boolean nature; with the exception of the command-line embedding, which is a 16-dimensional numerical vector. As a case in point, the dropped binary is represented by the following three security features, among others: the event type is \texttt{PROCESS}, the executable is located in the user directory \texttt{APPDATA}, and the binary is marked as dropped. As mentioned in Section~\ref{subsec:embedding-events}, the final embedding of all security features
is learned during the training phase of the neural network via back-propagation.

During the transformation of
a graph to a sequence, it is crucial to keep all security relevant information intact. Since the graph enrichment step uses the whole graph to add rich features to each node, the corresponding sequence represents the behavior of all processes in the provenance graph, and thus contains global information about the relationship between different events.

\textbf{Sequence to windows}. For efficiency, Transformers are typically trained with batches of fixed-size windows. Following this common practice, we slice each event sequence into fixed-size windows.
While windows only contain a limited number of \textit{local} events, they still contain \textit{global} information from the original provenance graph. This is due to the fact that graph enrichment happens on a global level (see Appendix~\ref{sec:appendix_global_graph_enrichment}).

For defining windows, we face an inherent trade-off between detection speed, which is required for a real-time detection system, and high malware detection accuracy. On one hand, it is imperative to use windows that are long enough, so that malicious patterns manifest themselves within one window. On the other hand, a detection system is supposed to be run on endpoints with minimal resource consumption, and therefore overly long windows are better avoided.
Our experiments show a significant gain in accuracy with increasing window size until around 200 (see Appendix~\ref{sec:appendix_window_size}). Beyond this, increasing the window size further does not lead to a significant accuracy improvement worth the additional detection latency; we therefore choose to use windows of 200 behavior events.

One more important design choice remains---the starting point of windows. A naive approach slices consecutive windows with neither gaps nor overlaps. 
However, we find that most malware, e.g., the Infostealers, execute malicious payloads soon after a new process has started. Averaged over both malicious and benign application samples of the REE-2023 dataset (Section~\ref{subsec:ree-2023}), process creations make up roughly 1\% of all observed behavior events, and they therefore offer a good candidate as
window starting points.

\vspace{-0.05cm}

\subsection{Implementation of \name}

\name consists of the data processing pipeline, which turns system logs into windows of security features, and the Transformer model for window classification.
The Transformer model for the REE-2023 dataset (Section~\ref{sec:dataset}) has a stack of four encoders, eight attention heads, and an internal token dimension of 60.
The Transformer is trained from scratch on labeled benign and malicious sequences to perform binary classification. 
Training is quick and takes 26 minutes
on an Nvidia GPU V100, thereby allowing for rapid hyperparameter tuning.
See Appendices~\ref{sec:appendix_training_progress} and~\ref{sec:appendix_hyper_params} for more details. The Transformer model is relatively small at a parameter size of 2.3 MB, and is therefore suitable for lightweight inference (see Section~\ref{subsec:resources} for real-time resource measurements).
The implementation of the Transformer model in \name as well as the data processing pipeline are available on our GitHub repository. The model, implemented using the TensorFlow library~\cite{tensorflow2015-whitepaper},
takes only \(\sim \)200 lines of code.

%% file: perf-eval.tex
\section{Performance evaluation}
\label{sec:evaluation}

We describe the datasets used for evaluation (Section~\ref{sec:dataset}), followed by introducing the state-of-the-art malware detection solutions that we compare against (Section~\ref{subsec:baselines}).
The comparison of \name with these baselines is done in Sections~\ref{subsec:baselines-results-ree} and~\ref{subsec:baselines-results-darpa}.
We then analyze~i)~the capability of the Transformer model (Section~\ref{subsubsec:transformer-model-results}), ii)~the benefit of command-line embeddings in \name (Section~\ref{subse:command-line-results}), and iii)~the impact of the security features (Section~\ref{subsec:sec-features-results}).

\vspace{-0.05cm}

\subsection{Datasets}
\label{sec:dataset}

We use two datasets for evaluations; see  Table~\ref{tab:dataset_size} for an overview. 

\vspace{-0.05cm}

\subsubsection{Real-world Enterprise EDR (REE-2023) dataset}
\label{subsec:ree-2023}
This proprietary dataset contains data which we collected from an enterprise environment.
To monitor the behavior of both benign and malicious applications, we leverage a commercial EDR tool, which uses ETW~\cite{etw} and Sysinternals~\cite{sysinternals} to monitor kernel calls.
The complete dataset is 15 GB in size and contains 29.2 million behavior events and 
12,700 provenance graphs. The benign part of the dataset was gathered during four months 
in 2022 and 2023, from four different enterprise users located in three different countries in Asia and Europe. The dataset contains behavior from hundreds of applications running on Windows OS, used by the users while carrying out their daily work.
For the malicious part of the dataset, 
we deployed a safe sandbox environment to execute malware and collect the corresponding data, thus ensuring that no damage is inflicted by malware on real endpoints. All malware samples were downloaded from VirusTotal in 2022 and 2023, are in executable format for Windows, 
and were executed on a special CAPE sandbox. We worked together with security engineers to tune the
CAPE sandbox, to
make it hard for the malware to detect the sandbox environment. 
We performed sanity checks to see if the samples included malicious activities, and that they were not malfunctioning. 
Since the automatic start of malicious samples in the sandboxing environment always happens in the same way, we do not include the first process start in the provenance graph.
Lest class-related artifacts are left in the data, we apply the same logic to benign provenance graphs.
On average, the benign and malicious provenance graphs of the REE-2023 dataset have a size of 2,300 behavior events, and the graphs have an average depth of 3.5 nodes. 
For the REE-2023 dataset, we use four event types, namely: new process creations, file access operations, network connections, and Windows registry events. For the benefit of the research community, we release the malware dataset via our GitHub repository. However, the benign dataset is private and has sensitive user information; for ethical reasons, we do not release the benign dataset. 

\paragraph*{Ethical concerns} The benign data was collected from the corporate computers of employees used for regular work. This was a voluntary program, and interested users were informed of the details of data collection---what will be collected, the purpose, and the period the data will be retained. No personally identifiable information (PII) of users was stored; any user names in the collected data (e.g., file names) were anonymized before storing. The data is stored securely on a company server, where only a limited group of authorized employees have access to.
The retention policy states that the data will be deleted six months after the research phase is completed. 

\begin{table}[]
\caption{Overview of datasets}\label{tab:dataset_size}
\vspace{-0.2cm}
\begin{center}
\begin{tabular}{lll}
\toprule
\textbf{REE-2023 dataset} & \textbf{Benign} & \textbf{Malicious} \\
\midrule
Graphs & 5,580 & 7,160 \\
Events & 6.14 million & 23.1 million \\
Windows of 200 events & 9,140 & 9,140 \\
\midrule
\textbf{DARPA-5D dataset} & \textbf{Benign} & \textbf{Malicious} \\
\midrule
Graphs & 721 & 91 \\
Events & 130,000 & 22,800 \\
Windows of 200 events & 2,370 & 2,520 \\
\bottomrule
\end{tabular}
\end{center}
\vspace{-0.5cm}
\end{table}

\subsubsection{DARPA FIVEDIRECTIONS (DARPA-5D) dataset}
The Transparent Computing (TC) program of DARPA released multiple public datasets~\cite{darpa_tc} containing system logs from APT attacks (advanced persistent threat).
The TC program held engagements number~3 and~5, in 2018 and 2019, respectively. Both engagements consisted of a red team trying to penetrate a target network and exfiltrate sensitive information. Since attacks were carried out manually, the malicious parts of both datasets are limited in size and variability, and we therefore choose to combine both datasets to form one larger dataset.
We focus on the attacks tracked by the FIVEDIRECTION team; they targeted hosts running Windows~10.
The red team used a variety of attack vectors, including Firefox backdoors and phishing emails with malicious MS Office macros.
Labeling of the provenance graphs (as malicious and benign) requires manual work, as the public version of the DARPA TC dataset has no labels attached to behavior events. We use the same labeled dataset as ProvNinja~\cite{mukherjee2023evading}, which includes the following behavior events: file creations, process creations, and network socket creations. Details on how we preprocessed and labeled this dataset can be found in Appendix~\ref{sec:appendix_darpa_3d}.

\vspace{-0.1cm}

\subsection{Baselines for evaluation}
\label{subsec:baselines}

We compare \name with two existing malware detection systems: ProvDetector~\cite{wang2020you} and DeepCASE~\cite{van2022deepcase}.

\textbf{ProvDetector}~\cite{wang2020you}
is proposed as an anomaly detection solution that learns to distinguish between common system entity interactions and rare ones. This frequency map is  used to find rare paths in a provenance graph, where a series of rare events are connected together. Subsequently, a document embedding~\cite{le2014distributed} is learned on the rare paths. At test time, anomalies in the embedding space are detected as malware. 
To make a fair comparison with \name, we adapt ProvDetector to perform supervised classification. 
We do so by replacing the final anomaly detector model with a Random Forest classifier. In this revised setting, ProvDetector learns from both malicious and benign graphs. 
We contacted the authors, and they helped us re-implement their closed-source solution.
As an additional reference, we leveraged an open-source re-implementation of ProvDetector by Goyal et~al.~\cite{goyal2023sometimes}.
To avoid overfitting to the training data, we use the same abstraction method as in the original paper, and remove user names and root directories from paths of files and executables, and mask out local IP addresses.

We carried out extensive hyperparameter tuning to achieve the best possible results with ProvDetector. On the REE-2023 dataset, we obtain the best validation results with a doc2vec dimension of 100. For classification of path embeddings, we tried both SVMs (support vector machines) and Random Forest classifiers, where the latter model achieved better results.
Since the classifier should produce a probability score
(so as to obtain the ROC curve), we use the percentage of malicious paths to compute a prediction probability per graph. 

Additionally, our analysis reveals that there is large variability in the file paths and process names in REE-2023 dataset. Since ProvDetector relies on the fact that system entities in the training data have the exact same name as in the test data, we apply several measures to remove any adverse effect of the datasets. First, we remove alphanumerical identifiers from file and process names. Second, we remove paths, since they vary significantly between users, and only keep file or process names. Finally, we use 100 rare paths per graph during training, instead of the 20 rare paths proposed in the original paper. These steps helped in obtaining the best results with ProvDetector on the REE-2023 dataset.

\textbf{DeepCASE}~\cite{van2022deepcase} is a semi-supervised system
which aims at reducing the workload for operators of a SIEM (security information and event management) system.
DeepCASE makes use of a GRU with attention model, which is trained to take in a sequence of security events and predict the next event. The attention-weighted GRU states are extracted and passed to a clustering algorithm. An analyst finally reviews a few samples per cluster, to perform labeling in semi-supervised way. Since our goal is to have a fully automated security solution, we use the labels of event sequences, i.e., ground truth, to automatically assign a label to a cluster, thereby turning it into a supervised solution. 
We use the open-source implementation~\cite{deepcase_code} 
of DeepCASE for our experiments. 

For achieving good results, we perform several adjustments to the DeepCASE implementation. First, we replace the SGD solver with Adam~\cite{kingma2014adam}. Next, we use probabilistic cluster labels, instead of binary ones: malware probabilities are based on the \textit{average} number of malicious samples in a given cluster. Finally, to avoid overfitting, we add weight regularization, in addition to the existing dropout layer. 
The hyperparameters for DeepCASE are provided in Appendix~\ref{sec:appendix_hyper_params}. We optimized the configuration for each experiment to achieve the best result.
The sequence length for DeepCASE is set to the same as that of \name, that is 200 events; note, this value is much higher than the 10 events used in the original paper~\cite{van2022deepcase}.

\subsection{Results}

For comparison of different solutions on a given dataset, we always use the same training:validation:test split of 60:20:20. We perform hyperparameter tuning for each experiment separately, on the validation dataset; see Appendix~\ref{sec:appendix_hyper_params} for details. A model's accuracy is reported on the test dataset.
Evidently, data seen during training does not appear during testing.

\textit{Metrics:} For performance evaluations, we use the common metrics of TPR (true positive rate) and FRP (false positive rate), where malware is the positive sample. 
In practice, the false positives, i.e., benign behavior being classified as malicious, are a burden to the security analysts who have to go through the alerts manually, leading to {\em alert fatigue}.
Therefore, it is important to measure the TPR at low values of FPR.
To demonstrate the trade-off between low FPR and high TPR, we plot the ROC (receiver operating characteristic) curves for all experiments.
We also report the AUC (area under the curve)
and the classification accuracy (percentage of correct predictions made).  Note, \name and DeepCASE~\cite{van2022deepcase} are sequence classifiers. Whereas, ProvDetector~\cite{wang2020you} works on whole provenance graphs; therefore, we report ProvDetector results for graph classification.

\vspace{-0.05cm}

\subsubsection{Comparison of \name with  baselines, on REE-2023 dataset}
\label{subsec:baselines-results-ree}

We now compare \name with the state-of-the-art solutions, namely DeepCASE~\cite{van2022deepcase} and ProvDetector~\cite{wang2020you}.
In real-world scenarios, endpoints will be hosting both well-known applications as well as some rare or new applications. For this reason, security solutions should have the capability to learn from existing benign applications, and generalize this behavior to \textit{unseen} applications. To emulate such a scenario, we split the benign part of the REE-2023 dataset by {\em end user}. Therefore, the test dataset contains behavior data from users whose data is not present in training.

\begin{figure*}[h]
    \centering
    \subfloat[\centering \name vs. baselines\newline on REE-2023]{\label{fig:roc_curves_baselines}\includegraphics[scale=0.675]{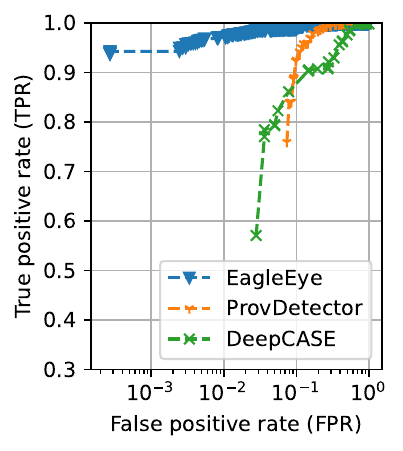}}
    \subfloat[\centering \name vs. baselines\newline on DARPA-5D]{\label{fig:roc_curves_darpa}\includegraphics[scale=0.675]{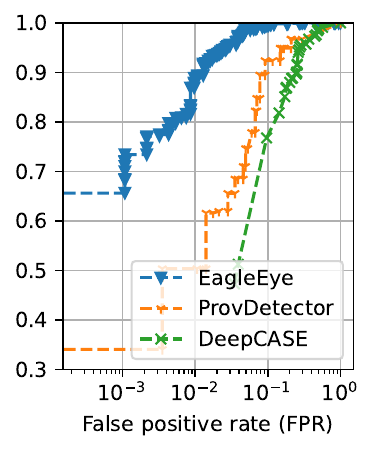}}
    \subfloat[\centering Comparing different\newline ML models]{\label{fig:roc_curves_ml_models}\includegraphics[scale=0.675]{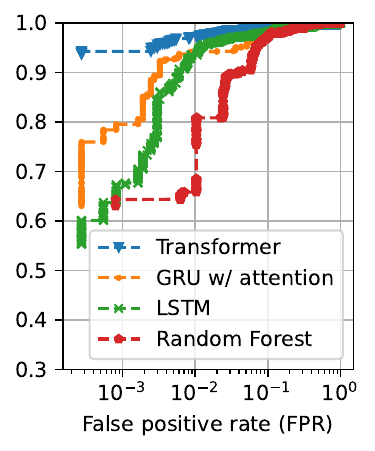}}
    \subfloat[\centering Command-line embeddings \newline \& security features]{\label{fig:roc_curves_security_features}\includegraphics[scale=0.675]{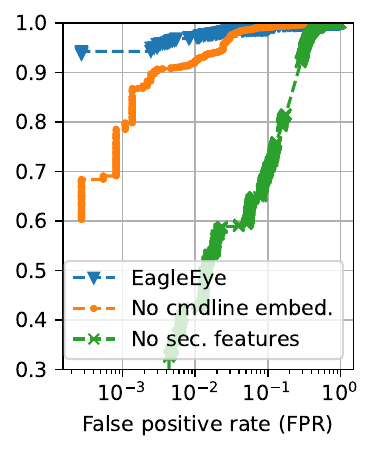}}
    \caption{ROC curves for malware detection (with FPR in logarithmic scale)} 
    \label{fig:rocs}
\end{figure*}

Figure~\ref{fig:roc_curves_baselines} plots the ROC curves of the three solutions, and Table~\ref{tab:results_ours_baselines} summarizes the key performance metrics. 
\name performs better than both the baselines. Specifically, \name detects more than 97\% of all malware samples at a low FPR of $10^{-2}$. Importantly, \name
maintains a high TPR of 94.3\% even at a much lower FPR of $10^{-3}$. In comparison, ProvDetector shows good detection capability only until $\approx10^{-1}$ FPR, at which point a further reduction of false positives leads to a sharp decline in malware detection capability.

We note that the REE-2023 dataset contains hundreds of application types, whereas the original ProvDetector dataset contains only 23 applications.
The fact that system entities have varying names on different systems, and that new applications use new file and process names, explains why ProvDetector has limited classification capability on datasets with large variability. 

\begin{table}[]
\caption{\name vs. baselines, on REE-2023 dataset}
\vspace{-0.1cm}
\label{tab:results_ours_baselines}
\begin{center}
\begin{tabular}{llll}
\toprule
\textbf{} & \textbf{\name} & \textbf{ProvDetector} & \textbf{DeepCASE} \\
\midrule
AUC & 99.7\% & 94.5\% & 92.5\% \\
TPR at $10^{-1}$ FPR & 98.8\% & 92.4\% & 86.1\% \\
TPR at $10^{-2}$ FPR & 97.1\% & 0\% & 0\% \\
TPR at $10^{-3}$ FPR & 94.3\% & 0\% & 0\% \\
Accuracy & 94.8\% & 90.4\% & 79.7\% \\
\bottomrule
\end{tabular}
\end{center}
\vspace{-0.5cm}
\end{table}

DeepCASE demonstrates the weakest performance, achieving a test accuracy of only 80\%.
This solution uses coarse event features, namely event types,
which means it does not utilize
all available data from provenance graphs.
In addition, predicting the next behavior event turns out to be challenging for the DeepCASE model; in comparison to a GRU with attention, a Transformer works better on long sequences of log data (see results in Section~\ref{subsubsec:transformer-model-results}).

\subsubsection{Evaluation on DARPA-5D dataset}
\label{subsec:baselines-results-darpa}

Next, we evaluate the malware detection capability of \name vs. the baselines on the public DARPA-5D dataset. 
When compared to REE-2023, the DARPA-5D dataset has several
key differences.
DARPA-5D contains APT attacks, which span over a longer period of time, whereas the REE-2023 dataset contains 
mostly malware samples which execute their payload  quickly.
In other words, the malicious behavior is spread across longer time durations in DARPA-5D. 
Second, DARPA-5D is a highly imbalanced dataset, whereas REE-2023 contains a similar amount of benign and malicious data.
Leveraging all available system logs from DARPA-5D, we get only 91 \textit{malicious} graphs, but 12.3K \textit{benign} graphs (see Table~\ref{tab:dataset_size}).
To alleviate this imbalance,
we first undersample benign graphs. Second, we oversample malicious event sequences for the creation of windows.
In order to achieve this oversampling, we drop the requirement that each window must start with a process creation. Note that our oversampling strategy entails that windows can have overlapping events; however, no two windows are identical. After this data rebalancing, we get $\approx2.4$K benign and $\approx2.5$K malicious windows, to evaluate \name and DeepCASE. 
Oversampling of malicious data works slightly differently for ProvDetector, as the system works on provenance paths, rather than event windows. We duplicate malicious graphs in the following manner: we sample from each graph's rarest paths, randomly pick 70\% of the paths, and create a new graph from them. We repeat this process until the dataset is balanced.

As shown in Figure~\ref{fig:roc_curves_darpa}, \name outperforms the baselines; and between the baselines, ProvDetetecor again performs better than DeepCASE. At $10^{-2}$ FPR, \name achieves an accuracy of  $\approx89\%$, which is around 38\% more than the next best-performing solution---ProvDetector. Table~\ref{tab:results_darpa} presents the overall results.

\begin{table}[]
\caption{\name vs. baselines, on DARPA-5D dataset}
\vspace{-0.1cm}
\label{tab:results_darpa}
\begin{center}
\begin{tabular}{llll}
\toprule
\textbf{} & \textbf{\name} & \textbf{ProvDetector} & \textbf{DeepCASE} \\
\midrule
AUC & 99.5\% & 95.1\% & 92.4\% \\
TPR at $10^{-1}$ FPR & 99.5\% & 92.4\% & 76.8\% \\
TPR at $10^{-2}$ FPR & 88.9\% & 50.4\% & 0\% \\
TPR at $10^{-3}$ FPR & 65.6\% & 34.1\% & 0\% \\
Accuracy & 96.5\% & 84.8\% & 82.9\% \\
\bottomrule
\end{tabular}
\end{center}
\end{table}

\subsubsection{Importance of the Transformer model}
\label{subsubsec:transformer-model-results}

\begin{table}[]
\centering
\caption{Comparison of ML models, on REE-2023 dataset}
\vspace{-0.1cm}
\label{tab:results_ml_models}
\begin{center}
\begin{tabular}{lllll}
\toprule
\textbf{} & \textbf{Transformer} & \textbf{GRU} & \textbf{LSTM} & \textbf{RF} \\
\midrule
AUC & 99.7\% & 99.2\% & 99.2\% & 98.5\% \\
TPR at $10^{-1}$ FPR & 98.8\% & 98.2\% & 98.1\% & 97.2\% \\
TPR at $10^{-2}$ FPR & 97.1\% & 94.2\% & 93.5\% & 65.8\% \\
TPR at $10^{-3}$ FPR & 94.3\% & 79.5\% & 67.3\% & 64.3\% \\
Accuracy & 94.8\% & 94.2\% & 93.3\% & 79.4\% \\
\bottomrule
\end{tabular}
\end{center}
\vspace{-0.5cm}
\end{table}

As an ablation study, we keep the data processing pipeline of \name fixed, but use alternative ML models for sequence classification, namely:
a GRU with attention, an LSTM (without attention), and a Random Forest.
Note, LSTM models have been experimented with in previous works~\cite{deeplog-2017, alsaheel2021atlas, villarreal2021hunting}. For example, ATLAS~\cite{alsaheel2021atlas} trains an LSTM to classify a sequence as attack or non-attack.
For each experiment, we
keep the data input and features exactly the same---each model is trained on behavior events from the REE-2023 dataset, where an event is represented by the feature vector described in Section~\ref{sec:enrich_graphs}.
Figure~\ref{fig:roc_curves_ml_models} and Table~\ref{tab:results_ml_models} present the results.

\textbf{Importance of Transformer architecture}: Among the different ML models, the Transformer architecture performs the best. The advantage of the Transformer architecture becomes apparent when we observe the detection rate at low FPR. At a limit of $10^{-3}$ FPR, the Transformer-based \name system achieves a detection rate (TPR) of 94.3\%, which is significantly higher than the second-best model's TPR of 79.5\%.
This validates our hypothesis that the Transformer model offers the best capability to extract global patterns in sequence data. Both the
GRU and LSTM model have to process behavior events sequentially, whereas the Transformer can look at the whole behavior sequence in one time step.

\textbf{Importance of event order}: As a conventional ML model alternative, we perform an experiment using a Random Forest (RF) classifier. 
In this experiment, inputs are individual events, not sequences, and labels are taken from the sequence that an event comes from. Therefore, classification is performed on a bag of behavior events without considering the order. In terms of prediction accuracy, the RF model shows significantly lower performance, achieving only 79.4\% accuracy, as compared to 93.3\% or higher for sequence models. This illustrates that the order of events in a behavior sequence contains valuable information about the intent of an application.

\subsubsection{Impact of command-line embeddings}
\label{subse:command-line-results}

The embedding of command-line~strings
into provenance graphs, and subsequently to sequences, is a novel aspect of \name; therefore, we evaluate it independently in this section.
Recall, our security features used for representing process creations are rather coarse-grained (see Table~\ref{tab:sec_features}). It is for this reason that we add embeddings of command-line strings to the feature vector, so as to encode more nuanced information about process creations.
In this experiment, 
we train a Transformer on the sequences of the REE-2023 dataset, but drop all command-line embeddings from the feature vectors.
Figure~\ref{fig:roc_curves_security_features} and 
Table~\ref{tab:results_security_features} present the results. The
plots (with and without command-line embeddings) have an increasing gap in detection rate as the FPR decreases; at a low FPR of $10^{-3}$, the absolute TPR increase due to command-line embeddings is 15.8\%, demonstrating the value of command-line strings for malware detection. 

\begin{table}[]
\caption{Impact of command-line string embeddings and security features; one at a time}
\vspace{-0.1cm}
\label{tab:results_security_features}
\begin{center}
\begin{tabular}{llll}
\toprule
\textbf{} & \textbf{\name} & \textbf{No cmdline} & \textbf{No security} \\
 & & \textbf{embeddings} & \textbf{features} \\
\midrule
AUC & 99.7\% & 99.5\% & 91.9\% \\
TPR at $10^{-1}$ FPR & 98.8\% & 99.1\% & 70.3\% \\
TPR at $10^{-2}$ FPR & 97.1\% & 92.2\% & 42.4\% \\
TPR at $10^{-3}$ FPR & 94.3\% & 78.5\% & 15.7\% \\
Accuracy & 94.8\% & 93.5\% & 81.4\% \\
\bottomrule
\end{tabular}
\end{center}
\vspace{-0.5cm}
\end{table}

\subsubsection{Relevance of security features}
\label{subsec:sec-features-results}

As a final ablation study, we assess the benefit of using graph-based security features instead of raw features. While most existing works use raw features like event types or raw file and process names, we enrich graph nodes with a set of features that are relevant for security purposes (see Section~\ref{sec:enrich_graphs}). We perform an experiment where we keep the architecture of \name  fixed, but ignore all security features from the behavior sequences, and only keep the type of each behavior event. Results on the REE-2023 dataset are reported in Figure~\ref{fig:roc_curves_security_features} and Table~\ref{tab:results_security_features}. At an FPR of $10^{-3}$, the detection rate of \name drops from  94.8\% to an extremely low 15.7\%.
This supports the argument that feature engineering at the graph level is required to achieve good prediction accuracy.

\subsubsection{Resource requirements for real-time malware classification}
\label{subsec:resources}

We now measure the overhead incurred for malware detection, when the endpoint in question has an average number of user applications running.
We use a commodity laptop with an Intel 3.6 GHz i7~$11^{th}$~Gen with 8~cores, 32 GB of main memory, but no GPU.
We include all steps of our data pipeline in the measurements, from preprocessing of the behavior logs, enrichment of graphs, command-line embedding, one-hot encoding of features, to the final inference with the Transformer model.
We execute \name in the background on an endpoint that runs several computer applications concurrently.
For real-world conditions, we assume an average number of behavior events based on the REE-2023 dataset. 
Our results show that \name uses an average CPU load of 2.65\% and an average memory utilization of 533~MB, which is acceptable
for real-world deployments~\cite{PG-EDR-CCS-2023}.

We note two important points. i)~We assume that behavior events are monitored separately with standard EDR tools; thus we ignore this workload for our experiment. That said, modern EDR tools use various optimization techniques such as minifilter drivers~\cite{minifilter} and non-blocking messaging queues to keep resource requirements 
at a minimum.
ii)~In our \name implementation,
more than 80\% of compute time is spent on data normalization and one-hot encoding. Therefore, the data processing step offers room for performance optimization.

%% file: case-studies.tex
\section{Case study}
\label{sec:case-study}

\begin{figure}[h]
\vspace{-0.5cm}
  \centering
  \includegraphics[width=8.5cm, clip, trim=5.3cm 2cm 12.5cm 4.1cm]{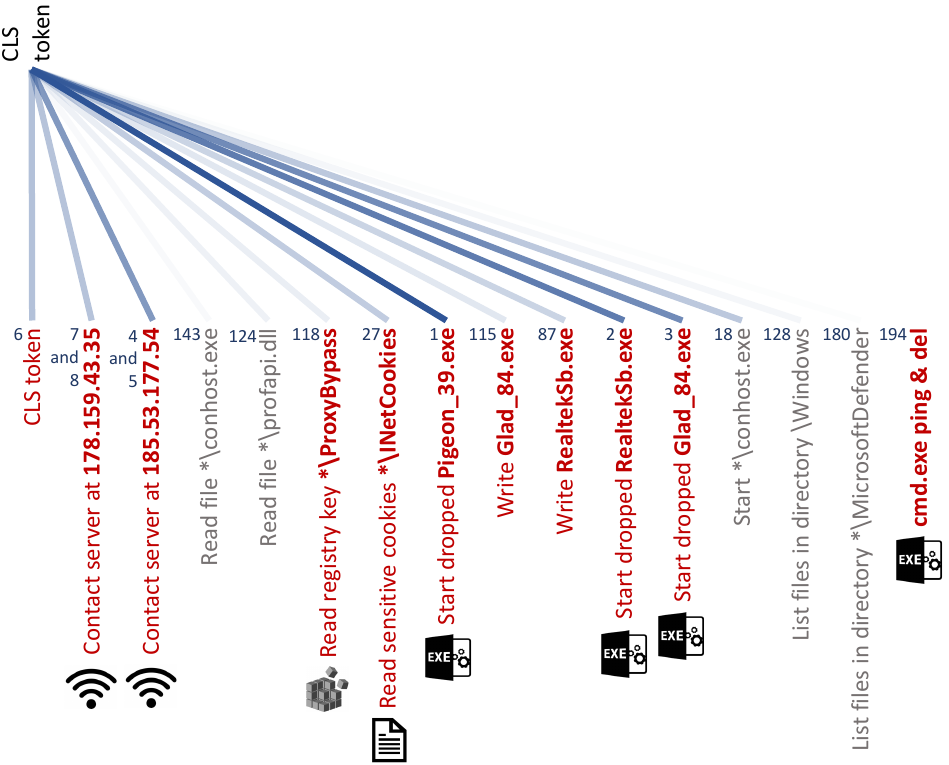}
  \vspace{-0.1cm}
  \caption{Attention ranking of \name's Transformer on Raccoon Infostealer behavior events. The lines connected to the CLS token are labeled with the attention ranks. Suspicious events are shown in red.}
  \vspace{-0.25cm}
  \label{fig:attention_scores}
\end{figure}

\name can not only detect malware at high accuracy, but it can also explain \textit{why} an application is malicious; in other words, it offers \textit{interpretability}. To achieve this goal, we leverage the Transformer model's attention mechanism, which learns to put the highest attention on the most suspicious behavior events.
As a case study, we revisit the Raccoon Infostealer (Section~\ref{sec:why_features}), which was not part of the training data. We proceed as follows. First, we query \name with the malware behavior sequence. Then,
we take an attention head from the last layer of the Transformer's attention stack (refer to Figure~\ref{fig:transformer_architecture}). Next, we measure the normalized attention scores from the representation of the CLS token, the rationale being that, \name's classification head applies only to the final CLS token representation (not the whole sequence).

We report the attention head scores in Figure~\ref{fig:attention_scores}. The figure presents the top 10 \textit{suspicious} events that got the highest attention scores, along with 5 randomly picked normal events, all ordered by time. 
Each event's attention is shown as a line to the CLS token, where the attention score is depicted via darkness of the line.

Observe from Figure~\ref{fig:attention_scores} that the starting of three dropped executables are the events most attended to. The next most important events are two outgoing Internet connections and the CLS token representation from the previous encoder layer. We performed a lookup on VirusTotal for the two IP addresses, which confirmed that they have been used as C\&C servers during the time. 
We note that each C\&C server is contacted via two distinct socket events (hence the corresponding lines are labeled with two ranks), but for readability, each event is only shown once. Although the last event, \texttt{cmd.exe} with self-deletion, is clearly a suspicious one, it 
did not get significant attention. 
A potential reason is that the REE-2023 malware dataset does not contain many samples where malware abuses existing executables on the target host. By adding more so-called Living-off-the-land examples to the dataset, the model for command-line embedding would potentially learn to better represent such commands.

We highlight that, based on the attention score, 7~of the top-8 highest-ranked events are indeed \textit{suspicious} events. The only exception is the CLS token representation, as should be expected.
Thus, presenting the top-10 events ranked by the attention scores from \name would help a security analyst to quickly interpret
a malware detection alarm; e.g., if most of the top-10 events look benign, the security analyst
can discard the result as a false positive.

As the Transformer model is queried with a sequence of 200 behavior events, its capability to detect the seven most malicious events is promising, thereby demonstrating the capability to extract malicious patterns across long sequences.

\vspace{-0.1cm}

%% file: related.tex
\section{Related work}

Over the years, cyber security systems have evolved from basic antivirus solutions
that search hashes of known malware, to sophisticated EDR solutions~\cite{EDR-Gartner} monitoring an extensive list of low-level system events, such as process creations, access to the file system, registry modifications, etc.
This fine-grained information is an advantage that provides the much-needed visibility to track process behavior~\cite{OmegaLog-NDSS-2020}.
Hence, a common approach for
malware detection on endpoints is to formulate behavior rules, which can then be matched against the collected system logs (e.g., see~\cite{slueth-2017, holmes-2019, Rapsheet-2020, gen-rules-USENIX-SEC-2023}). However, the logs also pose a challenge due to the vast scale of the collected data, motivating the need for ML models.
There exist multiple works that train anomaly detection models on log data~\cite{yuan2020ada, deeplog-2017, meng2019loganomaly,sierra-2022}. For example, DeepLog learns the behavior of events from recorded history of an endpoint and predicts future events with the goal of detecting anomalies~\cite{deeplog-2017}. 

An important recent advancement is the representation of
system events via
provenance graphs, thereby helping to capture and present the inherent relationships between the different events~\cite{OmegaLog-NDSS-2020, PG-EDR-CCS-2023}. These visual graphs are useful not only for analysts in investigating and interpreting application behavior, but also for developing new security solutions.
Based on provenance graphs, recent works looked into providing a high-level abstraction of the low-level events~\cite{NoDoze-2019, Rapsheet-2020, zeng2021watson}, investigating incidents of threats and attacks~\cite{alsaheel2021atlas, van2022deepcase}, and  detecting anomalies~\cite{manzoor2016fast, NoDoze-2019, wang2020you, threatrace-2022,ProGrapher-2023}. 

Broadly, existing works have the following limitations: either the information they extract as features is
limited, or the ML algorithm used to train a model from rich and dependent events falls short in learning the context of suspicious events.
Earlier solutions, e.g., StreamSpot~\cite{manzoor2016fast} and Unicorn~\cite{han2020unicorn}, deconstruct the graph into smaller structures of a fixed size, using embedding functions like StreamHash~\cite{manzoor2016fast} and HistoSketch~\cite{histosketch-2017}, to subsequently cluster the embeddings using appropriate distance functions for detecting anomalies. As illustrated in Section~\ref{sec:why_features}, encoding only graph structure is insufficient for attack detection, and evasion becomes easy~\cite{goyal2023sometimes, mukherjee2023evading}.
On the other hand, solutions like Sierra~\cite{sierra-2022} do not capture sequential dependency of behavior of events; besides, by using only network logs coming from firewall and IDS systems, protection systems lack the nuanced low-level information that would be available at endpoints.

ProvDetector~\cite{wang2020you} finds rare sequences of events in provenance graphs and applies the doc2vec embedding on these paths. Our work uses a more advanced event embedding, which is {\em learned} during Transformer training and uses all available context, not just the adjacent behavior events. As shown in our experiments in Section~\ref{subsec:baselines-results-ree}, benign applications may contain rare file and process names. ProvDetector is based on the assumption that all anomalies are malicious, which leads to false alarms in the presence of new applications or changing user behavior.

Other research works have proposed sequence models that are better suited for learning application behavior or application patterns~\cite{villarreal2021hunting, deeplog-2017, alsaheel2021atlas, van2022deepcase}. 
DeepLog~\cite{deeplog-2017} trains an LSTM model to detect, both, execution path anomalies and parameter value anomalies. Atlas~\cite{alsaheel2021atlas}, developed for constructing attack patterns from a given graph, uses an LSTM model as well. The goal of this work is to aid in attack investigation, and therefore it relies on a threat alert from another source.
DeepCASE uses GRU with attention, and trains the system in a semi-supervised way, to provide the context of behavior from event sequences. As observed in our experiments, DeepCASE has limitations in detecting malware with high accuracy, not only because it does not consider security-relevant fine-grained information (Sections~\ref{subsec:baselines-results-ree} and~\ref{subsec:baselines-results-darpa}), but also due to the inability of the GRU model to learn from long sequential data (Section~\ref{subsubsec:transformer-model-results}). A similar performance bottleneck is also observed with LSTM models (Section~\ref{subsubsec:transformer-model-results}). 

Recent works have started to explore GNNs (Graph Neural Networks)~\cite{han2021sigl, threatrace-2022, shadewatcher-2022}. For example, ShadeWatcher~\cite{shadewatcher-2022} proposes to use a GNN-based recommendation approach to detect malicious interacting entities (edges) on a
graph built from security logs. However, a graph in ShadeWatcher carries limited information, and consequently also does not provide contextual interpretation~\cite{argus-SP-2024}. Besides, solutions
which use the entire graph have
a detection lag as they have to wait until the graph is constructed. Similar challenges face ThreaTrace~\cite{threatrace-2022}. While promising, GNN models require
large datasets to learn effectively from sparse provenance graphs. 

%% file: conclusions.tex
\section{Conclusion}

To address the challenges in endpoint security, we present \name, a novel system based on a Transformer model. We demonstrate that \name can effectively learn the context of individual actions in long event sequences. Our solution uses rich security features, including
a representation of command-line strings which captures the
process tree hierarchy. Our evaluation on two datasets demonstrates \name's capability in achieving high detection rates at low FPR. Moreover, our system provides an explanation for malicious behavior via the attention scores of the Transformer model. 
We publish the implementation of \name and encourage researchers to use the source code as a starting point to make further enhancements. 

Given the potential of large language models (LLMs) in cyber security~\cite{divakaran-LLM-security-2024, lee-LLM-phishing-2024}, and the inherent language capability of LLMs, a promising next step is to integrate LLMs in endpoint security solutions. Foundational models for security~\cite{sharif2024drsec} promise to be adaptable to multiple downstream tasks (triaging, malware detection, etc.), while requiring fewer labeled data, and simultaneously providing explainability.

%% file: appendix.tex
\appendices

\section*{Appendix}

\section{Security features}
\label{sec:appendix_security_features}

\name uses a novel feature engineering approach to achieve superior malware detection performance. The \textit{raw} features of each action in the process provenance graph are used to compute structured \textit{security} features. Table \ref{tab:sec_features} presents a list of all security features used by \name.

\begin{table}[]
\small{
\caption{Security features defined in \name}
\label{tab:sec_features}
\begin{center}
\begin{tabular}{lll}
\toprule
\textbf{Event type} & \textbf{Security feature} & \textbf{Feature type} \\
\midrule
Process start & Name of executable & Categorical \\
 & Path of executable & Categorical \\
 & File extension of executable & Categorical \\
 & Dropped binary & Boolean \\
 & Length of command-line string & Numerical \\
 & Number of command-line flags & Numerical \\
 & Command-line embedding & Numerical vector \\
File access & File path & Categorical \\
 & File extension & Categorical \\
 & Access mode (w/r/d) & Categorical \\
 & File access options & Categorical \\
 & Sensitive file & Boolean \\
 & Access amount & Numerical \\
Registry & Internet key & Boolean \\
access & Persistence key & Boolean \\
 & Uninstall key & Boolean \\
 & Notify key & Boolean \\
 & Data type of key & Categorical \\
Network & Is source internal & Boolean \\
connection & Is destination internal & Boolean \\
 & Service port & Categorical \\
 & Connection size & Numerical \\
 & Transport layer protocol & Categorical \\
 & Incoming or outgoing & Boolean \\
All events & Time duration & Numerical \\ 
\bottomrule
\end{tabular}
\end{center}
}
\end{table}

\section{Trade-off between window size and accuracy}
\label{sec:appendix_window_size}

\begin{figure}[h]
  \centering
  \includegraphics[clip, trim=0cm 0cm 1cm 1cm, width=6cm]{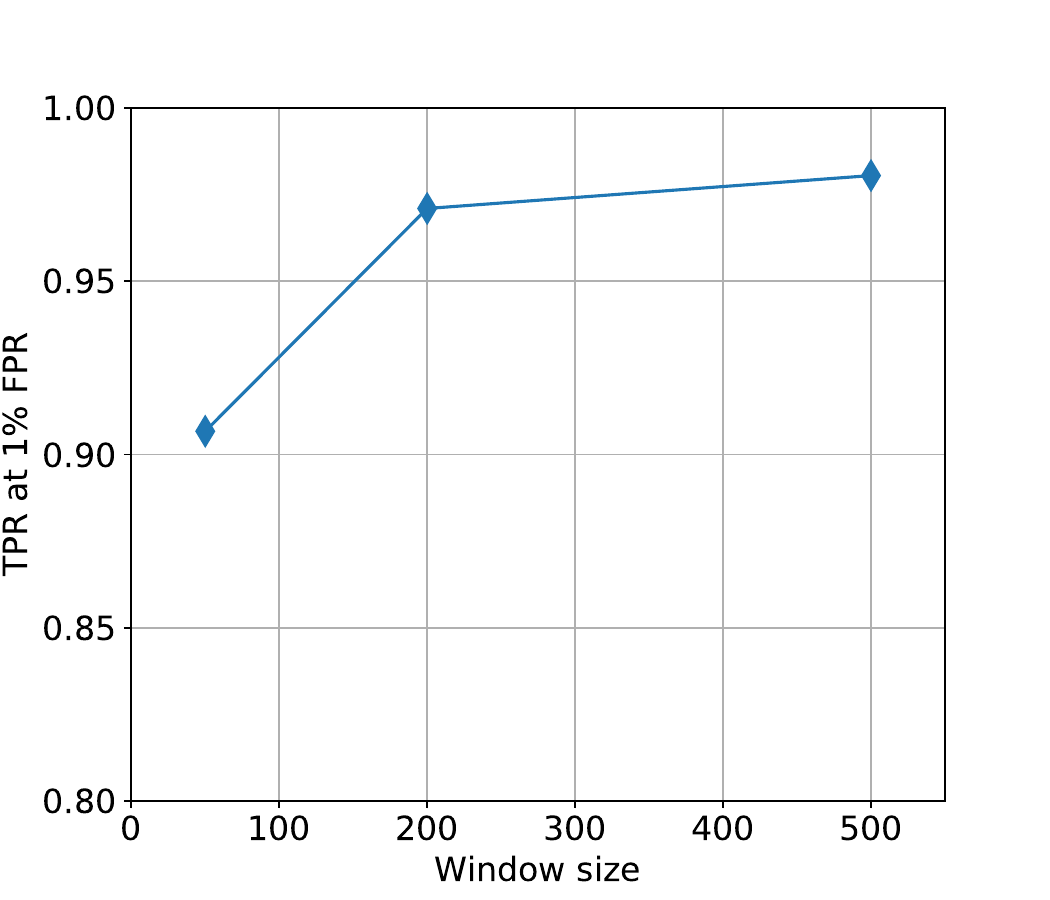}
  \caption{Trade-off between window size and classification accuracy}
  \label{fig:window_size_tradeoff}
\end{figure}

\name uses sequences of behavior events as its input. 
Longer sequences  contain more information and thus lead to better classification accuracy. However, short sequences offer the benefit of faster malware detection, as well as lower resource requirements for real-time malware protection on endpoints.

In Figure~\ref{fig:window_size_tradeoff}, we present the malware detection rate for different window sizes, specifically at an  FPR limit of $10^{-2}$. For a window size of 200, the TPR is 97.1\%. When we reduce the window size to 50, TPR drops to 90.7\%, which is a significant loss in detection capability. When we increase the window size from 200 to 500, we observe
a marginal TPR increase of only 1\%.
Therefore, we select a window size of 200 for our evaluation in Section~\ref{sec:evaluation}, which strikes a good balance between detection latency and classification accuracy.

\section{Training progress of \name's Transformer}
\label{sec:appendix_training_progress}

\begin{figure}[h]
  \centering
  \includegraphics[clip, trim=0cm 0cm 1cm 1cm, width=8cm]{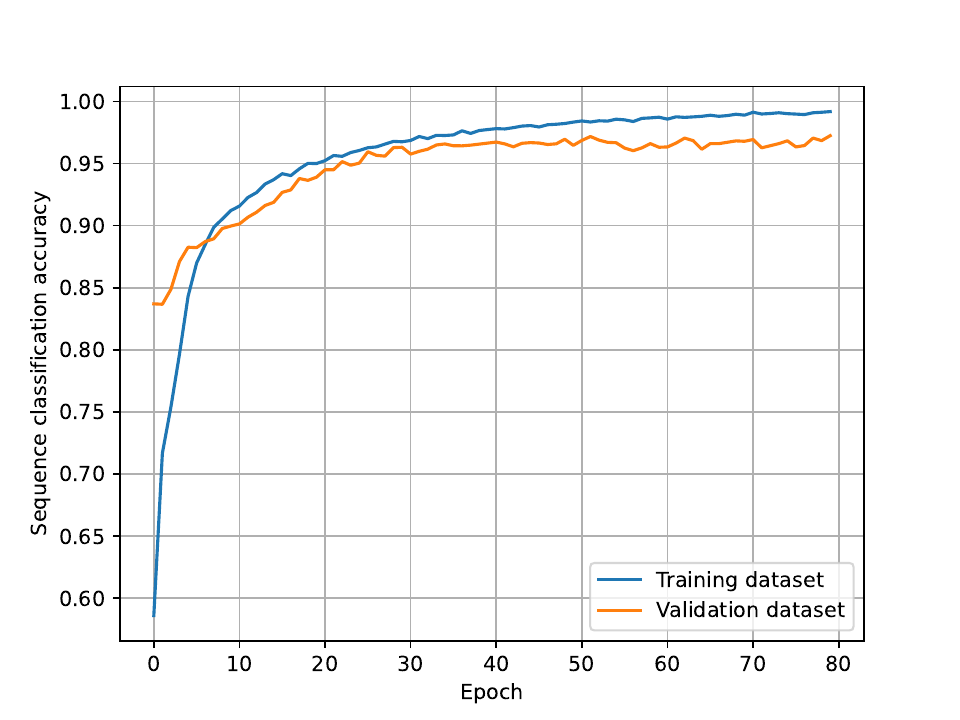}
  \caption{Training of \name's Transformer on the REE-2023 dataset}
  \label{fig:appendix_secformer_training}
\end{figure}

\name uses a Transformer model to detect malicious event sequences. The training progress of the best-performing model on the REE-2023 dataset is shown in Figure~\ref{fig:appendix_secformer_training}. The training accuracy continuously rises for the first 80 epochs. The validation accuracy lags slightly behind the training accuracy, which is probably due to a slightly more challenging validation dataset. While training accuracy increases till the end, the validation accuracy doesn't grow significantly after 40 epochs. Our hyperparameter tuning shows that a low learning rate is crucial for achieving a solution
that generalizes well to unseen data.

\section{Hyperparameters for machine learning models}
\label{sec:appendix_hyper_params}

In this section, we give more details about the hyperparameters of all the machine learning models used during our evaluation. We include architecture details for \name, as well as the baselines, ProvDetector~\cite{wang2020you}, and  DeepCASE~\cite{van2022deepcase}.

\begin{table}[]
\caption{Hyperparameters and model size for best \name Transformer}
\label{tab:appendix_hyper_params_secformer}
\begin{center}
\begin{tabular}{lll}
\toprule
 & REE-2023 & DARPA-5D \\
\midrule
Total parameter size & 2.3 MB & 1.3 MB \\
Training time for all epochs & 26 min & 28 min \\
Number of encoders & 4 & 6 \\
Number of attention heads & 8 & 6 \\
Token dimension & 60 & 40 \\
Training epochs & 80 & 140 \\
Learning rate & $10^{-5}$ & $5*10^{-5}$ \\
Dropout & 0.1 & 0.1 \\
Weight regularization & $10^{-2}$ & $10^{-2}$ \\
Dimension of attention keys & 60 & 40 \\
Dimension in FF & 240 & 160 \\
Batch size & 64 & 4 \\
\bottomrule
\end{tabular}
\end{center}
\end{table}

Table~\ref{tab:appendix_hyper_params_secformer} shows hyperparameters of \name's best Transformer model. The Transformer model takes in as input, a sequence of behavior events represented by security features, and returns a malware prediction probability. In the table, \textit{Token dimension} refers to the dimensionality of the vector used to represent the hidden state of each token in the encoder stack. \textit{Dimension in FF} refers to the dimension of the hidden layer in the feed-forward neural network, which is located between multi-head self-attention layers. L2 Weight regularization is applied to the dense layers in the classification head. Dropout is applied in several places: after token embedding, in the multi-head self-attention layers, in the feed-forward neural network, and in the classification head's dense layer.

\begin{table}[]
\caption{Hyperparameters and model size for best ProvDetector classifier}
\label{tab:appendix_hyper_params_provdetector}
\begin{center}
\begin{tabular}{lll}
\toprule
 & REE-2023 & DARPA-5D \\
\midrule
Rare paths (train) & 100 & 100 \\
Rare paths (test) & 20 & 20 \\
Maximal path length & 10 & 10 \\
Doc2vec model size & 238 MB & 3 MB \\
Doc2vec embedding dimension & 100 & 20 \\
Doc2vec training epochs & 50 & 100 \\
Random Forest model size & 5.3 GB & 1.37 GB \\
Random Forest trees & 2000 & 2000 \\
Maximal tree depth & 20 & 20 \\
\bottomrule
\end{tabular}
\end{center}
\end{table}

Next, Table~\ref{tab:appendix_hyper_params_provdetector} shows the hyperparameters for our best ProvDetector implementation, on both the REE-2023 dataset and DARPA-5D. We provide numbers for the doc2vec model, which is responsible for embedding rare sentences of varying length into a fixed-size vector representation. Moreover, we show details of the Random Forest model, which classifies the fixed-size vectors into benign and malicious. \textit{Maximal path length} refers to the maximal length for rare paths. Longer paths are cut into multiple parts.

\begin{table}[]
\caption{Hyperparameters and model size for best DeepCase classifier}
\label{tab:appendix_deepcase}
\begin{center}
\begin{tabular}{lll}
\toprule
 & REE-2023 & DARPA-5D \\
\midrule
Model size & 5.4 MB & 4.6 MB \\
Sequence length & 200 & 200 \\
Training epochs & 3 & 5 \\
GRU hidden nodes & 64 & 64 \\
Epsilon for DB Scan & $10^{-2}$ & $2*10^{-2}$ \\
Minimum cluster size & 5 & 3 \\
Weight regularization & $10^{-2}$ & $10^{-2}$ \\
Learning rate & $5*10^{-3}$ & $10^{-4}$ \\
Confidence threshold & 0 & 0 \\
Solver & Adam & Adam \\
Cluster labels & probabilistic & probabilistic \\
\bottomrule
\end{tabular}
\end{center}
\end{table}

Finally, Table~\ref{tab:appendix_deepcase} lists hyperparameters for the best DeepCASE models on REE-2023 and DARPA-5D. Weight regularization was applied to the GRU sequence model. The confidence threshold was set to zero, to
ensure we always get a prediction.

\section{DARPA-5D dataset}
\label{sec:appendix_darpa_3d}

The DARPA TC program released system log tracing data for engagements number 3 and 5. During engagement~3, attackers used Firefox to download and execute a backdoor named DRAKON as part of the APT group's toolset; moreover attackers used phishing emails with malicious MS Office macros to install malware beacons. For engagement~5, attackers used a Firefox backdoor to download and execute the payload DRAKON and MICRO backdoor.

Since the official DARPA dataset is not labeled at the event level, we describe here how we performed labeling, preprocessing of the data, and provenance graph creation. We thank the authors of ProvNinja~\cite{mukherjee2023evading} for providing us with a labeled version of the DARPA TC-3 and TC-5 dataset. For creation of labeled graphs, the publicly available red team reports are leveraged. With help of the reports, malicious system entities used during an attack are located, and the entities are used as starting points to grow malicious provenance graphs. The growing of the provenance graphs is done via breadth-first search with a maximal depth of~8. Malicious system entities can include malicious URLs, sensitive files, abused living-off-the-land binaries, and dropped binaries.

Benign behavior happens concurrently with attacks. Such behavior includes normal web browsing, use of Office applications to create documents, and reading of
emails.
Note that benign and malicious provenance graphs can overlap in time. Once malicious graphs are created, all remaining unused behavior events are leveraged for the creation of benign graphs. Attacks in the TC program start with either Firefox or Microsoft Excel. Both applications also get used regularly for benign activity. To
make the malware detection task as realistic as possible, both benign and malicious graphs contain at least one Firefox application; additionally, most graphs also contain a Microsoft Excel application.

Note, our labeled version of the dataset might differ slightly from other works, as separating benign from malicious activity is a manual process and entails various design decisions. In particular, our dataset consists of \textit{provenance graphs}, where each node has exactly one incoming edge, namely from its parent process.

\section{Enrich events with global context}
\label{sec:appendix_global_graph_enrichment}

\begin{figure}[h]
  \centering
  \includegraphics[clip, trim=4cm 1.1cm 15.9cm 11.8cm, width=8cm]{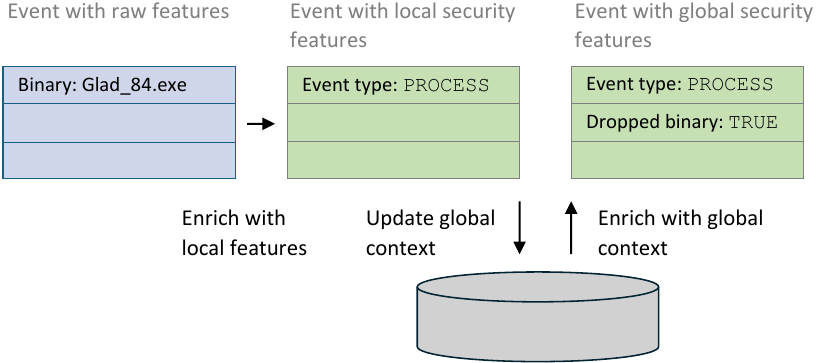}
  \caption{\name uses global context from the process provenance graph for computation of security features.}
  \label{fig:appendix_global_graph_enrichment}
\end{figure}

\name runs inference on windows of 200 events to detect malware. While event windows are of limited size, they first go through an enrichment phase and thus contain global information from the whole provenance graph.
A resource-efficient implementation necessitates a global context
which stores the process call hierarchy as a tree, as well as a list of persisted binaries.
This global data structure allows us to discard old behavior events after a grace period.

Figure~\ref{fig:appendix_global_graph_enrichment} depicts the use of both a local and global context. As a case in point, the figure shows a behavior event which represents the start of a binary called \texttt{Glad\_84.exe} from the previously discussed Raccoon Infostealer. For each event, \textit{local} raw features are used to compute \textit{local} security features. In this case, the system can infer the event type as \texttt{PROCESS}. Next, the global context is queried to infer global security features. The fact that this binary was dropped might not be visible in this or the neighboring behavior events. Instead, the global data structure contains the necessary information that a binary with exactly this name and path has been written to disk before. As a final step, the global context is updated with the raw information of this particular event. In this case, the call hierarchy tree is extended by this process. In the end, the event is represented by all security features, and now contains both local and global information.